\documentclass[10pt,letterpaper]{article}
\usepackage{opex3}

\usepackage{amsmath,amssymb,graphicx}

\newcommand{\LPoi}{LP$_{01}$}
\newcommand{\LPoii}{LP$_{02}$}

\begin{document}

\title{Experimental realization of femtosecond transverse mode conversion using optically induced transient long-period gratings}
\author{Tim Hellwig,$^{1,*}$ Martin Schnack,$^1$ Till Walbaum, Sven Dobner and Carsten Fallnich}
\address{Institute of Applied Physics, Westf\"alische Wilhelms-Universit\"at, Corrensstrasse 2, 48149 M\"unster, Germany\\$^1$Authors share credit in equal parts\\}
\email{$^*$tim.hellwig@wwu.de}

\begin{abstract}

We present the experimental realization of transverse mode conversion in an optical fiber via an optically induced long-period grating. The transient gratings are generated by femtosecond laser pulses, exploiting the Kerr effect to translate intensity patterns emerging from multimode interference into a spatial refractive index modulation. Since these modulations exist only while the pump beam is present, they can be used for optical switching of transverse modes. As only a localized part of the grating was written at a time and the probe beam was co-propagating with the pump beam the required pulse energies could be reduced to 120\,nJ which is about a factor of 600 lower than in previous quasi-continuous-wave experiments. Accompanying numerical simulations allow a better understanding of the involved effects and show excellent agreement to the experimental results.
\end{abstract}

\ocis{060.4370 (Nonlinear optics, fibers); 190.4420 (Nonlinear optics, transverse effects in); 190.3270 (Kerr effect); 190.4370 (Nonlinear optics, fibers).}

\section{Introduction}

In optical communication technology spatial-division multiplexing has gained a lot of attention to solve an expected capacity crunch in common communication schemes \cite{Richardson2013}. In order to further scale data capacities transverse eigenmodes of few-mode fibers could be used as multiplicator for having access to additional channels \cite{Carpenter2012,Amaya2013}. One challenging aspect of using transverse modes for data communication is the routing and switching purely by optical means. Conventionally, static periodic index perturbations of the fiber are used to achieve effective energy transfer between transverse modes \cite{Hill1990}. These long-period grating  structures inside the fiber are usually permanently written using UV \cite{Vengsarkar1996} or pulsed radiation \cite{Kondo1999a} or exploiting inherent photosensitivity in Germanium-doped fibers \cite{Park1989} and can therefore not easily be used for switching applications. However, transient gratings optically induced by counter-propagating high-power nanosecond pulses via the Kerr-effect (OLPG, optically induced long-period gratings) have recently been presented and used for transverse mode conversion \cite{Andermahr2010}. These optically induced gratings inherently offer applicability for optical switching, although the pulse energies used for these quasi-continuous-wave experiments were in the order of a few tens of micro-joule and thereby in the same order of magnitude as the damage threshold of the fiber facets.

\begin{figure*}[hbt]
\centering
\includegraphics[width=.7\textwidth]{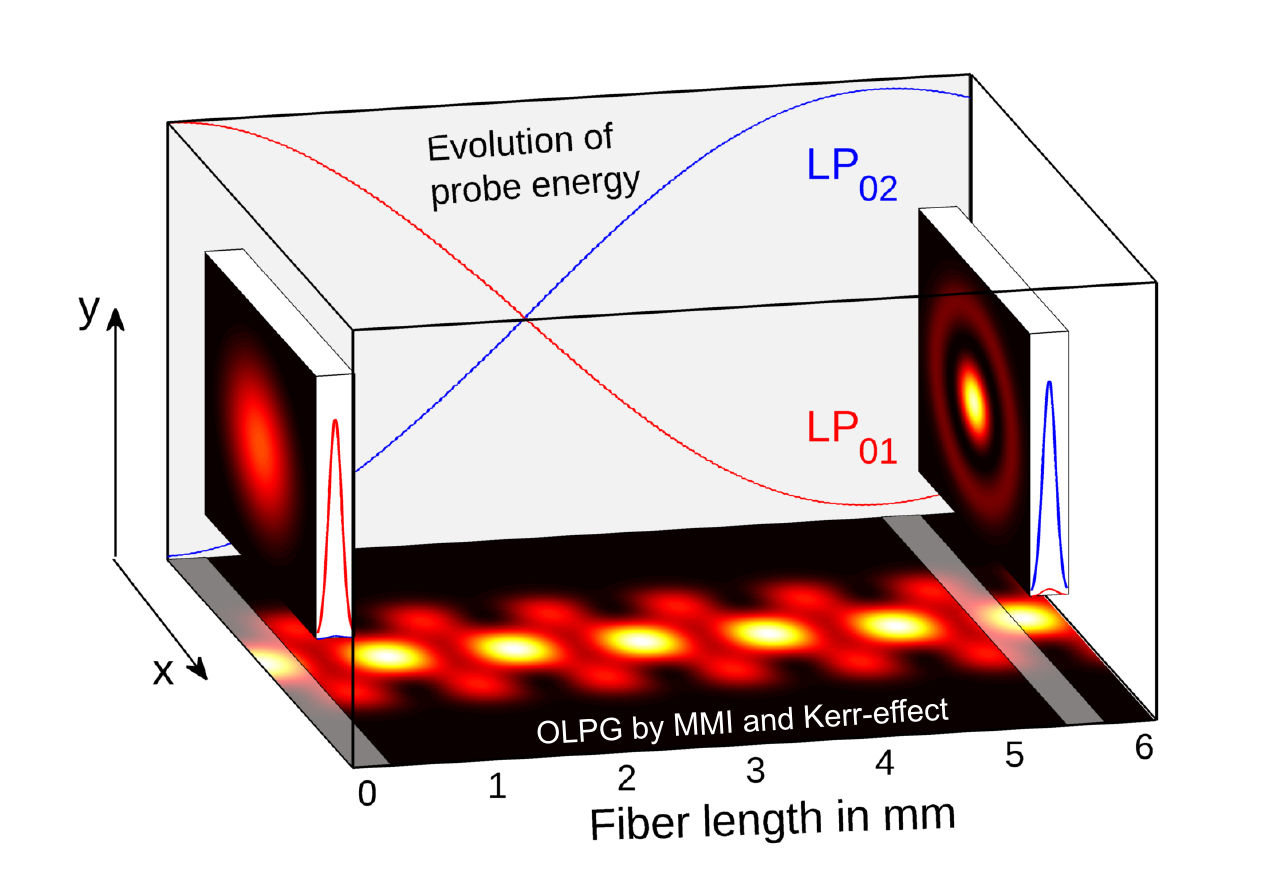}
\caption{Mode conversion by OLPGs: The intensity distribution of the multi-mode interference (MMI) of a corresponding cw-pump beam equally distributed between the \LPoi- and the \LPoii-mode is shown projected to the xz-plane while the change in modal probe energy is shown on the yz-plane, with z being the propagation direction. The probe pulse intensity distributions at the beginning of the fiber and after maximum conversion are shown in the insets as well as the temporal intensity profiles of both probe modes (red = \LPoi-, blue = \LPoii-mode). The localized parts of the OLPG induced by a femtosecond pump beam are indicated by the grey rectangles. The conversion speed is exaggerated in this schematic representation. See \textcolor{blue}{Media~1} for an animated version of the process.}
\label{scheme_fig}
\end{figure*}

In this paper, we present steps towards potential applications of OLPGs in telecommunications. In order to achieve the necessary index modulation for mode conversion in a fused-silica fiber the required peak power is fixed in the kilo-watt regime and, therefore, to reduce the required pulse energies to a feasible level (sub-nano-joule), only part of the grating can be induced at a time. This implies that the grating has to travel alongside the signal that is to be switched into a different mode. The use of co-propagating femtosecond pulses for the creation of an OLPG and the conversion of transverse fiber modes presented here, thereby enables to reduce the required pulse energy by almost three orders of magnitude (about a factor of 600) compared to \cite{Andermahr2010}. By coupling a pump beam into a mixture of two transverse modes (e.g., \LPoi\ and \LPoii) of an optical few-mode fiber, the differing propagation constants of the modes lead to multimode interference (MMI) along the fiber, creating the periodic spatio-temporal intensity pattern depicted in Fig.~\ref{scheme_fig} due to the shape of the modes. As the Kerr-effect links the refractive index to the local intensity, this results in a long-period phase grating, with the grating constant being defined by the difference of the propagation constants of the pump modes \cite{Walbaum2013}. However, only a part of the grating is induced at a given time with its length proportional to the temporal duration of the pump pulse (see sketch in Fig.~\ref{scheme_fig} and \textcolor{blue}{Media~1}). For the experiments a co-propagating and orthogonally polarized femtosecond probe beam at the same wavelength was coupled into the fundamental mode of the fiber. As the propagation constant of that probe beam then differed from the one of the \LPoii-mode by exactly the grating constant, the phase-matching condition for efficient energy transfer between the probe modes \cite{Bures2009} was inherently fulfilled. Thus, the probe beam was diffracted at the OLPG and conversion into the \LPoii-mode occurred while the probe beam traveled alongside the induced grating (see schematic representation of the conversion process in Fig.~\ref{scheme_fig}). One can calculate a rough estimate of the expected fundamental properties of the induced grating following coupled-mode theory \cite{Bures2009} by assuming a transversely uniform long-period grating with a cosine modulation given by the maximum induced refractive index change, e.g. $\delta$n$ \approx 2.0\cdot 10^{-5}$ for a 400\,fs Gaussian shaped pulse with 120\,nJ pulse energy. The resulting grating induced in the fiber studied here has a grating period of about $100\,\mu$m and a coupling length of $z_c \approx 5$\,cm. The full width half-maximum (FWHM) phase matching bandwidth calculated from the fiber's dispersion profile is then about 17\,THz, being more than 5 times broader than the bandwidth of the used pulses.  

\section{Experimental Setup}

\begin{figure*}[htb]
\centering
\includegraphics[width=.8\textwidth]{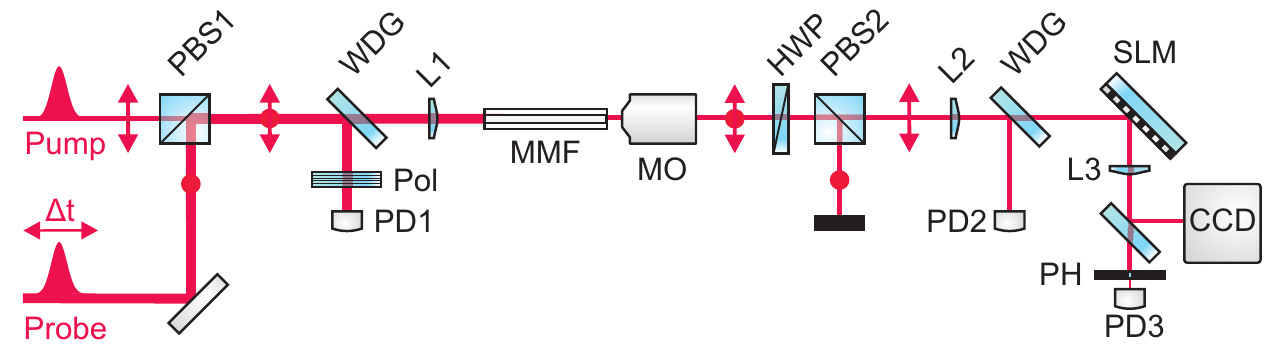}
\caption{Schematic diagram of the experimental setup (angles and distances are not drawn to scale). PBS: polarizing beam splitter, PD1-PD3: photo diodes, WDG: wedged glass substrate, L1-L3: lenses, MMF: multi-mode fiber, MO: microscope objective, HWP: half-wave plate, Pol: polarizer, SLM: spatial-light modulator, CCD: charged-coupled device camera, PH: pinhole. For further details see text.}
\label{setup_fig}
\end{figure*}

The experimental setup is shown in detail in \mbox{Fig. \ref{setup_fig}}: Pump (p-polarized) and probe beam (s-polarized) were provided by an amplified, mode-locked Ytterbium fiber laser delivering ultrashort pulses with a duration of approximately \mbox{400\,fs} at a repetition frequency of \mbox{1\,MHz} and a center wavelength of about 1033\,nm with a spectral width of about 11\,nm, corresponding to 3.1\,THz at the laser's central wavelength. The probe beam path also included a piezo-driven delay stage ($\Delta$t) for adjusting the temporal overlap between pump and probe pulses. After recombining the cross-polarized pump and probe beams with a polarizing beam splitter (PBS1) both were coupled into a few-mode fiber (nLIGHT Passive $25\,\mu$m core width, NA 0.08, 6.5\,cm fiber length). The diameter of the pump beam was chosen to be smaller than that of the fundamental fiber mode in order to excite a mixture of the \LPoi- and the \LPoii-mode and care was taken to maximize the temporal overlap of pump and probe pulses. The differential group delay of the involved modes in this fiber was calculated to be $24\,$fs/cm, corresponding to a delay of 162\,fs at the end of the fiber.  Although, this means that the grating strength did decrease along the propagation path, there was interaction between the fiber modes along the whole fiber length.

As the discrimination between probe and pump beam is based on polarization, any polarization distortion in the fiber, e.g., by nonlinear polarization rotation (NPR) leads to a cross-talk when measuring the signal behind the fiber. NPR strongly depends on the local polarization ellipticity in the fiber, and thereby on the phase difference of the cross-polarized beams at the entrance fiber facet. A small portion of both beams was extracted in front of the fiber for phase difference measurements and the following experiments were evaluated at a phase difference ensuring minimal NPR (for details see Section~\ref{sec:NPR}).

\begin{figure}[htb]
\centering
\includegraphics[width=.6\columnwidth]{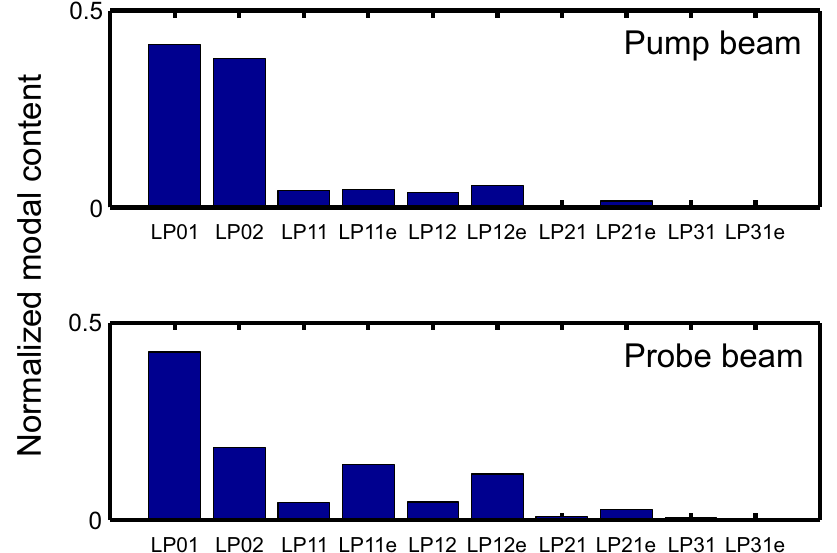}
\caption{Measured normalized excited modal contents  (orientational degeneracy marked with ``e'') in the Passive 25 fiber for the pump and probe beam individually.}
\label{modal_contents}
\end{figure}

The modal contents were measured with a modified version of the correlation filter presented by Flamm et al. \cite{Flamm2012}. The fiber end-facet was magnified by a 100x microscope objective and the beam of interest was filtered by means of a half-wave plate and PBS2.  The selected beam was then relay imaged to a phase-only spatial light modulator (SLM). By applying the SLM with pre-calculated phase patterns specific to each mode a power proportional to the modal weight was diffracted into the center of the first diffraction order in the far-field \cite{Flamm2012}. In comparison to Flamm et al. \cite{Flamm2012} we measured the optical power in the center of the diffraction order with a fast photodiode (PD) and a pinhole with a diameter of 10\,$\mu$m,  accurately aligned in front of the PD, instead of using a single pixel of a CCD camera. We then determined the position of the pinhole in the Fourier plane by applying the SLM with a phase grating and scanning its period. This modification allowed us to measure changes in the individual modal contents with a measurement bandwidth only limited by the bandwidth of the applied PD, and thereby to resolve the modal contents as a function of the relative phase between probe and pump beam. The feasibility of the modal reconstruction with the acquired values from the PD was verified by a separate measurement utilizing a CCD camera. Figure \ref{modal_contents} shows the measured normalized modal contents of both beams individually, when the transmission of the beam under test at PBS2 was maximized using the half-wave plate and the other beam was blocked. The pump beam had an almost equal distribution between the fundamental mode and the \LPoii-mode with only marginal power in other higher-order transverse modes. It was verified that the modal content of the pump beam did not change during the course of the mode conversion measurements by repeating the modal content measurement of the pump beam directly after each experiment. The maximum amount of energy of the probe beam was measured to be in the fundamental mode ensuring a good contrast for conversion measurements. As we were able to quantitatively measure changes in the modal contents, the excitation of some higher order modes in the probe beam due to a slightly elliptical beam profile was found to be unproblematic in the experiments.

\section{Cross-talk by nonlinear polarization rotation}
\label{sec:NPR}

In order to measure mode conversion in the probe beam, the s-polarized pump beam was suppressed by minimizing its transmission at PBS2. The contrast at PBS2 was good enough (about 100:1) so that linear interference between the transmitted residual pump beam and the probe beam was negligible when the probe beam was set to 10\% of the pump beam's power throughout the experiments. However, at the peak powers necessary for inducing a long period grating via the Kerr effect, the aforementioned nonlinear polarization rotation (NPR) occurred. In fact, NPR rotates the polarization ellipse of the input light field by an angle $\theta \propto \sin(\Delta\phi)$, where $\Delta\phi$ is the phase difference between the cross-polarized probe and pump beams. This rotation of the polarization ellipse effectively leads to a cross-talk between both beams and possibly disturbs the measurement:
\begin{figure}[hbt]
\centering
\includegraphics[width=.65\columnwidth]{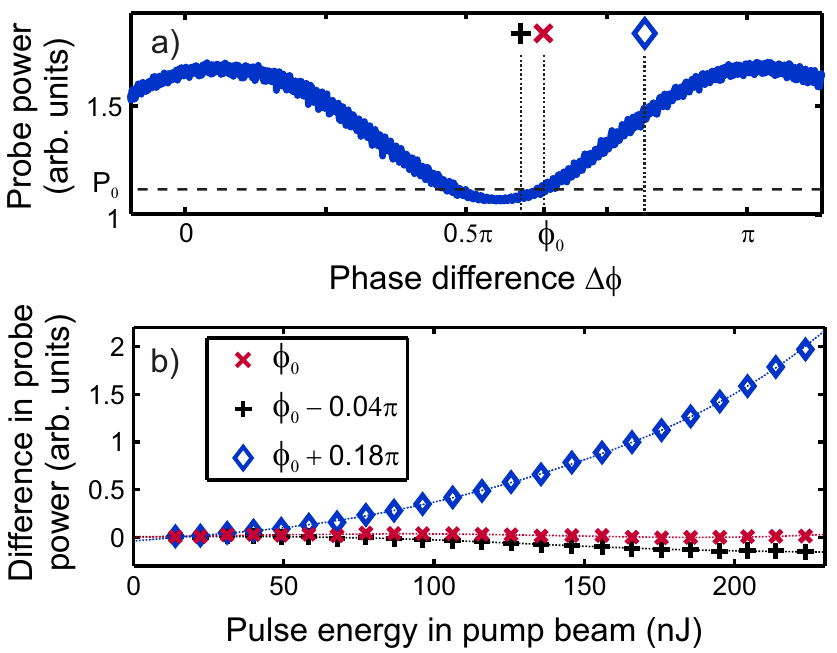}
\caption{\textbf{(a)} Probe power measured with PD2 as a function of the phase difference between probe and pump beam in front of the fiber at an average pump beam power of 150\,mW and an average probe beam power of 15\,mW. The horizontal dashed line indicates the probe power $P_0$ without a pump beam present. The markers indicate phase delays that are investigated in more detail in sub-figure \textbf{(b):}
Here, the pump energy dependent difference between the probe power levels ``with'' and ``without'' pump beam present is depicted for different phase differences measured with PD1 (dotted lines to guide the eye). For details see the text.}
\label{phase_fig}
\end{figure}
Due to unavoidable mechanical fluctuations in the setup, this phase difference varied over time and furthermore, even small contributions to the phase difference between both beams, caused by residual fiber birefringence, led to strong NPR. In order to be able to account for or even avoid this unwanted, obstructing cross-talk and thereby to be able to measure mode conversion, we studied the occurring NPR as a function of the phase difference. We measured the transmission of the probe beam's power through PBS2 behind the fiber (with PD2) while ramping the delay of the probe beam ($\Delta$t) with a sub-Hz frequency and simultaneously measuring the phase difference to the pump beam in front of the fiber with PD1. One result is depicted in Fig.~\ref{phase_fig}(a) for a pump pulse energy of 150\,nJ. The measured probe power shows a $\pi$-periodicity as it is expected from NPR. Corresponding to the modulation of the probe power the polarization ellipse, formed by the interference of pump and probe beam, was periodically rotated into and out of the probe beam's (p-)polarization direction. The average probe power $P_0$ measured with a blocked pump beam was therefore periodically exceeded and undershot. However, note that NPR only occurs for elliptically polarized light so that no NPR would have been expected at a phase difference of zero or $\pi$ in front of the fiber. The large difference in measured probe power from $P_0$ at these phase difference values could only be attributed to a phase shift due to residual birefringence in the fiber and showed that it cannot be neglected in the case of NPR. To identify the optimal input phase difference $\phi_0$ for minimal cross-talk between both beams, we measured the phase-dependent average probe output power for pulse energies ranging from 10\,nJ to 250\,nJ equivalent to measured average powers of 10\,mW to 250\,mW. We then calculated for each phase difference the sum of the root mean square (rms) deviations from the unaltered probe power values $P_0$ measured with a blocked pump beam. The phase difference at which the minimum rms deviation occured  is identified as $\phi_0$ in Fig.~\ref{phase_fig}a. The deviation from $P_0$ as a function of the pump beam pulse energy at $\Delta\phi=\phi_0$ is shown in detail in Fig.~\ref{phase_fig}(b). Furthermore, the power difference is depicted for $\Delta\phi=\phi_0+0.18\pi$ as well as for $\Delta\phi=\phi_0-0.04\pi$ to demonstrate the nonlinear transfer of pump power into the probe beam as well as of probe power into the pump beam. The former effect appears much stronger as the pump beam holds ten times the power of the probe beam. 

\section{Mode conversion}

\begin{figure}[hbt]
\centering
\includegraphics[width=.7\columnwidth]{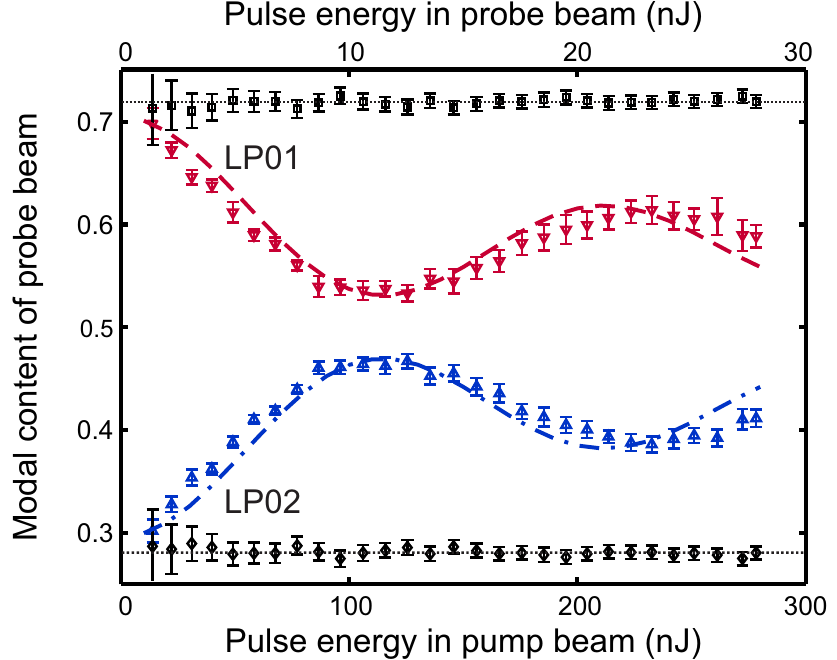}
\caption{Relative modal distribution between \LPoi- and \LPoii-mode as a function of the pump and probe pulse energy when the pump beam was blocked (black squares and diamonds) as well as when pump and probe beam overlapped in time (red and blue triangles). The results of a corresponding numerical simulation are presented as the red dashed and the blue dash-dotted curves (for details see text). A good contrast between the probe and the residual pump beam was required for accurate modal content measurements. As the polarization contrast at PBS2 was limited, the probe beam energy (upper x-axis) was therefore always set to 10\% of the pump energy (lower x-axis) during the experiment as well as within the simulation to be well above the residual pump beam energy.}
\label{conv_fig}
\end{figure}

We then measured the \LPoi- as well as \LPoii-modal content in the probe beam as a function of the pump beam's power and thereby as a function of the degree of index modulation of the OLPG. For each pump beam power the modal contents were measured as a function of the phase difference between both beams using photodiode PD3. The amplified PD3 utilized for modal power measurements had a measurement bandwidth of 12\,kHz well above the used phase-modulation frequency. The measurement was then evaluated at the optimal phase difference for minimal NPR, and the resulting normalized evolution of the \LPoi- as well as \LPoii-modal content is shown in Fig.~\ref{conv_fig}.

 The black squares and diamonds depict the modal content of the probe modes when the pump beam was blocked and therefore no OLPG was present. The relative modal content was constant within the measurement error. The same was found to hold true for an unblocked but delayed pump beam. In this case, although an OLPG was transiently induced, the instantaneous nature of the Kerr-effect led to no conversion if the probe pulses did not overlap in time with the only transiently induced grating. These tests also verified that no temperature-induced contribution to the OLPG was present in our setup. The red triangles in Fig.~\ref{conv_fig} finally show the evolution of the fundamental \LPoi-mode when the probe pulse overlapped in time with the OLPG while the blue triangles depict the evolution of the higher-order  \LPoii-mode as a function of the pump pulse energy. A conversion of probe energy from the \LPoi- to the \LPoii-mode could be observed, reaching its maximum conversion with a relative \LPoii / \LPoi-mode content of 46\% at a pulse energy of 120\,nJ equivalent to a measured average power of 120\,mW. For even higher pump beam pulse energies back-conversion occurred until the probe energy in the \LPoii-mode reached a local minimum at about 230\,nJ of pump pulse energy.

In numerical studies of femtosecond mode-conversion scenarios a conversion efficiency of about 77\% for identical probe and pump pulse lengths as used in our experiments, and more than 90\% for longer pump than probe pulses, have already been demonstrated \cite{Walbaum2013,Hellwig2013}. In order to get a better understanding why the experimentally observed mode-conversion efficiency differed from the numerical results the involved ultrashort pulses were studied in more detail: It was identified that the intensity auto-correlation of the pulses in use for the experiments revealed significant and unavoidable energy in a structure surrounding the 400\,fs pulse up to delays of about 5\,ps. As the pump and probe pulses were derived from the same source, this pedestal did affect the measured mode conversion in a twofold manner: Not only was the peak power of the pump pulse considerably reduced compared to bandwidth limited pulses, but also did the energy in the picosecond pedestal of the probe pulse not take part in the nonlinear conversion process due to not overlapping with the induced grating. However, the unaltered modal distribution of these temporal parts of the probe beam was still included in the average power values measured with PD3 for modal reconstruction, and thereby constituted an offset in the relative modal content, reducing the maximum achievable conversion efficiency. Therefore, we performed numerical simulations based on coupled nonlinear Schroedinger type equations (see \cite{Walbaum2013} for details on the simulations) specifically for the scenario studied here (dashed-dotted red and dashed blue curve in Fig.~\ref{conv_fig}). To account for the effective loss in peak power for mode conversion we assumed bandwidth-limited 400\,fs pulses but with reduced pulse energy. The linear offset in the final modal content was considered by adding after the simulation the missing probe pulse energy but with the initial, unaltered modal distribution. With these simple assumptions we observed very good agreement with the experimentally observed mode conversion (compare measured data with dashed-dotted red and dashed blue simulation curves in  Fig.~\ref{conv_fig}) when reasonably assuming  35\% of the pulse energy to be in the picosecond pedestal of the pump and probe pulses.  To get rid of the remaining small residual deviations between simulation and experiment it would be required to fully characterize phase as well as amplitude of the laser pulses in order to perform numerical simulations considering the complete pulse information.

\section{Conclusion}
In conclusion, we have shown femtosecond transverse mode conversion in a fiber by use of an optically induced long-period grating that travels along the fiber with the group velocity of the pump pulses. This grating could be switched on and off since it was transiently induced using the Kerr-effect, and relatively low pulse energies of \mbox{120\,nJ} in the pump beam were sufficient to induce the grating. It was demonstrated that mode-conversion could be measured despite a strong cross-talk between pump and probe beam via nonlinear polarization rotation by measuring the modal contents as a function of the phase difference between pump and probe.  The experiments were implemented in a proof of principle style by converting modal energy from the fundamental \LPoi- to the \LPoii-mode, as an equal mixture of these modes for the pump beam was easily excitable in the experiments. A relative  \LPoii / \LPoi-mode content of about 46\% in the probe beam after conversion was achieved experimentally. Mode-conversion between other transverse-mode pairs could be achieved by utilizing, e.g., phase plates or an additional spatial light modulator for excitation of the pump modes. 

The demonstrated reduction in needed pulse energy for inducing the grating was already almost three orders of magnitude (about a factor of 600) compared to preceding quasi-continuous-wave experiments \cite{Andermahr2010}. However, the necessary pulse energies need to be reduced further into the pJ-regime for this scheme to be applicable for all-optical routing in future spatial-division multiplexing communication systems. The according progress is expected to be achievable by the use of highly nonlinear waveguides, e.g. fibers made of chalcogenide glass or integrated waveguides made of silicon exhibiting an increased non-linearity by a factor of 100 up to 50000 \cite{Eggleton2011,Ding2014}. Furthermore, numerical simulations of multi-mode coupled nonlinear Schroedinger equations proved to be helpful for understanding the involved mechanisms, especially under the influence of non ideally shaped pump pulses, and showed excellent agreement with the experimentally measured mode-conversion. We will therefore continue to use numerical simulations parallel to experiments for the identification of the best suited nonlinear platform for all-optical mode switching. Currently we are eliminating the possible cross-talk of pump and probe beam originating from nonlinear polarization rotation by exploiting a dichroic setup that we already numerically studied elsewhere \cite{Hellwig2013,Schaferling2011}.

\section*{Acknowledgment}
We acknowledge support by the Deutsche Forschungsgemeinschaft and the Open Access Publication Fond of the University of Muenster.

\end{document}